\documentclass[preprint,showpacs,preprintnumbers,amsmath,amssymb]{revtex4}

\newcommand\br{\begin{eqnarray}}
\newcommand\er{\end{eqnarray}}
\newcommand\be{\begin{equation}}
\newcommand\ee{\end{equation}}

\newcommand\lb{\lbrack}
\newcommand\rb{\rbrack}

\newcommand\bc{\begin{center}}
\newcommand\ec{\end{center}}

\newcommand{\ct}[1]{\cite{#1}}
\newcommand{\bib}[1]{\bibitem{#1}}
\newcommand\NPB[3]{\textsl{Nucl. Phys.} \textbf{B#1}, #3 (#2)}

\newcommand\PRD[3]{\textsl{Phys. Rev.} \textbf{D#1}, #3 (#2)}

\newcommand\PLB[3]{\textsl{Phys. Lett.} \textbf{#1B}, #3 (#2)}

\newcommand\MPLA[3]{\textsl{Mod. Phys. Lett.} \textbf{A#1}, #3 (#2)}

\begin{document}

\title {Extra Dimensional Curvature  Supression of the Effective Four Dimensional Vacuum Energy Density }

\author{E.I. Guendelman}%
\email{guendel@bgumail.bgu.ac.il}
\affiliation{%
Department of Physics, Ben-Gurion University of the Negev \\
P.O.Box 653, IL-84105 ~Beer-Sheva, Israel
}%

\begin{abstract}
Considering a very large number of extra dimensions,  $N\rightarrow \infty$, we show that in the effective four dimensional picture, to leading order in $N$, both the cosmological constant in $N+4$ dimensions and the curvature of the extra dimensions (curved as spheres) give the same type of contributions. Furthermore in this limit, the extra dimensional curvature naturally supress the effect of a positive Cosmological Constant, so that the resulting effective potential governing the vacuum energy in the effective $4-D$ picture has a leading $1/N$ dependence (i.e. vanishing in the large $N$ limit). We can understand qualitatively this effect in a heuristic picture, by thinking that all dimensions, both visible and extra have an equal sharing of the curvature caused by $\Lambda$, in this case when increasing the overall number of dimensions by adding $N$ extra dimensions, then if $N$ is large, the visible dimensions do not have to curve too much, hence a small four dimensional vacuum energy follows. In the large $N$ picture the potential can be also stabilized by a small (i.e. vanishing at large $N$)  expectaction value of a four index field strength.
\end{abstract}

\pacs{04.50.Cd, 04.50.Kd}

\maketitle

\section{Introduction}

Many interesting ideas for approaching the unification of fundamental forces involve considering extra dimensions\ct{review}. The consideration of extra dimensions also affects our understanding of the cosmological constant problem, for example recently Canfora, Giacomini and Zerwekh \ct{CGZ} have considered the limit of large extra dimensions in General Relativity with a cosmological constant in $4+N$ dimensions, $N$ dimensions being compactified into a sphere. These solutions display the interesting property that the ratio of the four dimensional cosmological constant to the $4+N$ cosmological constant goes to zero as $N\rightarrow \infty$.

Those solutions obtained in \ct{CGZ} and constructed just from pure gravity plus cosmological constant are unstable however as we will see. Nevertheless, we show here that the $N\rightarrow \infty$ can be exploited in any case to explore the question of the smallness of the four dimensional cosmological constant. In fact, we will show that in the effective four dimensional picture, to leading order in $N$, both the cosmological constant in $N+4$ dimensions and the curvature of the extra dimensions give the same type of contributions, therefore in this limit extra dimensional curvature can naturally suppress  the effective $4-D$ Cosmological Constant. Furthermore in the large $N$ picture the instability can be cured and true ground states with a small vacuum energy can be obtained by enlarging the theory so as to include matter fields, like four index field strengths,  then the potential can be also stabilized by a small (i.e. vanishing at large $N$)  expectaction value of a four index field strength.
\

\section{Effective Four Dimensional Picture of Einstein Gravity Plus $N+4$ Cosmological Constant}
The dynamics of the system is then assumed to be defined by the very simple action
\begin{equation}\label{e1}
S = \int{\frac{1}{16\pi G}(R-2\Lambda)}\sqrt{-g} d^{4+N}x 
\end{equation}
\noindent
where $g =  det (g_{AB})$. $A, B = 0, 1, 2,.....,N+3$.
we now consider the case where the extra dimensions are compactified, but the size of the extra dimension can depend locally on the four dimensional location $x^{\mu}$, $\mu,\nu  =0, 1, 2, 3$, so we consider the metric
\begin{equation}\label{e2}
ds^{2} = g_{\mu\nu}(x^{\mu})dx^{\mu}dx^{\nu} + a^2 (x^{\mu})\frac{dy^jdy^j}{(1+\frac{1}{4}ky^2)^2}
\end{equation}
$j=4,....N+3$ and $y^2=y^jy^j$. The extra dimensions are compactified on a sphere ($k>0$)
and the $4D$ part of the metric $g_{\mu\nu}(x^{\mu})$ being $y^j$ independent.

Setting units where $8\pi G =1$, we observe that the equations of motion can be put in the form of the four dimensional general relativistic form with "matter", if we define, following for example \ct{DG}, where additional matter fields (four index field strengths) will not be considered at first (but in the following section  they will be shown to play a role in the stabilization of the compactified solutions).
\begin{equation}\label{e3}
\bar{g}_{\mu\nu}= a^N  g_{\mu\nu}
\end{equation}
\noindent
then we get
\begin{equation}\label{e4}
\bar{R}_{\mu\nu}-\frac{1}{2}\bar{g}_{\mu\nu}\bar{R} = - T_{\mu\nu}
\end{equation}
$T_{\mu\nu}$ being the energy momentum tensor of a $4D$ scalar field
\begin{equation}\label{e5}
\phi = \sqrt{\frac{1}{2}N(N+2)} lna
\end{equation}
\noindent
subject to the very special potential
\begin{equation}\label{e6}
V(\phi) = -\frac{1}{2}k N(N-1)exp(-2\sqrt{\frac{N+2}{2N}} \phi)+ 
\Lambda exp(-\sqrt{\frac{2N}{2+N}} \phi)
\end{equation}

Notice now that the limit $N \rightarrow \infty$  is very special, indeed, in this limit 
$2\sqrt{\frac{N+2}{2N}}=\sqrt{\frac{2N}{2+N}}=\sqrt{2}$. So to leading order the two terms in eq. (\ref{e6}) become both proportional to $exp(-\sqrt{2}\phi)$. In principle, the effect of a very large cosmological constant $\Lambda$ can be cancelled by  appropriately  curving the extra dimensions. The cancellation, for the choice $\Lambda=\frac{1}{2}k N(N-1)$ is exact only in the limit $N \rightarrow \infty$ .
For $N$ large but finite, the condition $\Lambda=\frac{1}{2}k N(N-1)$ can be achieved by a shift of the scalar field $\phi$, then for $N$ large but finite, subleading terms survive, but this is typical of a "renormalization" procedure: a counterterm is not expected to cancell a complete contribution (here of the cosmological constant), but only a leading part of it, we will now see how this works in details. 

For finite $N$ and for $k>0$ and $\Lambda>0$, thre is always a value of the scalar field $\phi$ such that the $k$ and $\Lambda$ contributions exactly cancel each other, for convenience, we want to consider this point as our origin. To do  this let us consider now for finite $N$ a shift the field  $\phi$ by a constant,
\begin{equation}\label{e7a}
\phi \rightarrow \phi + \Delta
\end{equation}
\noindent
in this case $\Lambda$ and $k$  transform as
\begin{equation}\label{e7b}
 \Lambda \rightarrow \Lambda exp(-\sqrt{\frac{2N}{2+N}} \Delta)
\end{equation}
and 
\begin{equation}\label{e7c}
 k \rightarrow k \\exp(-2\sqrt{\frac{N+2}{2N}} \Delta)
\end{equation}

Then for finite $N$ the condition $\Lambda=\frac{1}{2}k N(N-1)$ can implemented for an appropriate choice of $\Delta$, which for the case of large (but finite) $N$ is given by $\Delta = \frac{N}{2\sqrt{2}}ln (\frac{k N(N-1)}{2\Lambda})$\\this just means that we take conventionally the zero of the potential at $\phi = 0 $.  To see this, notice that $\Lambda=\frac{1}{2}k N(N-1)$ means that $V$ is now given by
\begin{equation}\label{onlyLambda}
V(\phi) = \Lambda \Bigl\lb  exp(-\sqrt{\frac{2N}{2+N}} \phi)-exp(-2\sqrt{\frac{N+2}{2N}} \phi)\Bigr\rb 
\end{equation}

\noindent
which can be expanded to first order in $1/N$ giving
\begin{equation}\label{e8}
V_{sl}(\phi) = \Lambda\\ exp(-\sqrt{2}\phi) \frac{2\sqrt{2}\phi}{N}
\end{equation}
\noindent
which displays explicitly a $1/N$ supression and the choice $V(0)=0$, but notice $\phi=0$ is not a minimum, but very close to being flat (if $N$ is very, very large). The above potential does not have any true ground state, just a maximum, no stable compactified state therefore. We discuss now the issue of stabilization and the possibility of obtaining a true ground state with small vacuum energy density, for this we cannot consider just pure gravity with a cosmological constant, at least at the classical level,  there is the need to introduce extra matter fields.

\section{Stabilization of the potential by adding a four index field strength}

The stabilization of the potential can be achieved for example by adding a four index field strength defined from a $3$
index potential $F_{BCDE} =\partial_{[B} A_{CDE]}$. These terms are present for example in supergravity theories 
\ct{CJS}, although there  no bare cosmological constant can be considered and it is specific to eleven dimensions while here we take a large number of dimensions. This is added to the action, giving now a total action of the form 

\begin{equation}\label{newaction}
S = \int{\Bigl\lb 
\frac{1}{16\pi G}(R-2\Lambda) - \frac{1}{48} F_{BCDE}F^{BCDE}}\Bigr\rb \sqrt{-g} d^{4+N}x 
\end{equation}
The equation of motion for $F^{BCDE}$ is

\begin{equation}\label{fourfield eq.}
\partial_B(\sqrt{-g} F^{BCDE}) = 0
\end{equation}
\noindent
there is a solution taking an expectation value in the visible four dimensional space (indices $\mu, \nu, \alpha, \beta$ below taking values $0, 1,2,3$ only),
 
\begin{equation}\label{fourfieldsol.}
 F_{\mu\nu\alpha\beta} = \Omega (x^\mu) \epsilon_{\mu\nu\alpha\beta}
\end{equation}
\noindent
$\epsilon_{\mu\nu\alpha\beta}$ being the Levi-Civita tensor in $4-D$, 
then $\Omega (x^\mu)$ in (\ref{fourfieldsol.}) is given by 

\begin{equation}\label{Omega}
 \Omega(x^\mu) = \frac{\lambda\sqrt{-g_{(4)}}}{3a^N}
\end{equation}
\noindent
this is similar to the Freund Rubin solution \ct{FR}, except that here we consider an explicit cosmological constant in higher dimensions. Here $g_{(4)}$ is the determinant of the four dimensional part of the metric, this then modifies the potential $V$, so we have to add a $\lambda$ dependent part to obtain, for the potential of $\phi$, defined  before (once again using units where $8\pi G =1$), and again working in the Einstein frame, we obtain
\cite{DG}

\begin{equation}\label{potpluslambdaterm}
V(\phi) = -\frac{1}{2}k N(N-1)exp(-2\sqrt{\frac{N+2}{2N}} \phi)+ 
\Lambda exp(-\sqrt{\frac{2N}{2+N}} \phi) + \lambda^2 exp(-3\sqrt{\frac{2N}{2+N}} \phi)
\end{equation}
proceeding with the choice  $\Lambda=\frac{1}{2}k N(N-1)$ and once again going to the $N \rightarrow \infty $ limit and
keeping the leading term in the $1/N$ expansion,
we obtain the potential, 
\begin{equation}\label{newpot}
V(\phi) = \lambda^2 exp(-3\sqrt{2}\phi) + \Lambda\\ exp(-\sqrt{2}\phi) \frac{2\sqrt{2}\phi}{N}
\end{equation}
\noindent
the new $\lambda^2$ term can be used to stabilize the potential, for example, we can achieve a minimum at $\phi =0$, by requiring the absence of linear terms in $\phi$, when Taylor expanding (\ref{newpot}) near $\phi =0$,
this requirement gives
\begin{equation}\label{sollambd}
 \lambda^2 = \frac{2\Lambda}{3N}
\end{equation}
\noindent
and for this choice, the next term in the Taylor expansion, that is the quadratic contributions in $\phi$ for $V$, add up to give the positive (therefore indicating stability) contribution $\frac{2\Lambda \phi^2}{N}> 0$ if $\Lambda > 0$. Finally, for this example, the value of $V$ at the minimum is now $V= \lambda^2 = \frac{2\Lambda}{3N} > 0$, if $\Lambda > 0$, giving rise therefore to a small (for $N$ large) positive vacuum energy density in the effective four dimensional picture.

Choices that differ slightly from (\ref{sollambd}) will shift slightly the location of the minimum and the value of the vacuum energy density at that minimum.
\section{Conclusions}
We have seen that for large number of extra dimensions, the effect of the curvature and the effects of the higher dimensional cosmological constant become degenerate, the extra dimensional curvature can act in this limit as an effective counterterm for the cosmological constant term. A Planck scale Cosmological constant could be then reduced by appropriately  curving the extra dimensions in this  limit, may be a new way to look at the cosmological constant problem.

Notice that here even before we concern ourselves with the value of the potential at its extremum, we notice that the $1/N$ supression is shown here to be valid for the functional form of  the potential, not just for the potential at some special point.  However playing with just curvature of extra imensions and a cosmological constant in higher dimensions does not lead to a stable ground state. The "ground state" in \cite{CGZ}, is not stable, since it represents the maximum of the potential, but the $1/N$ suppression claimed in \cite{CGZ} is a genuine effect, not because of the value of the vacuum energy at this particular extremum, but because the potential is suppressed everywhere in this limit.  

We can understand qualitatively the  effect explained above in a heuristic picture, by thinking that all dimensions, both visible and extra have an equal sharing of the curvature caused by $\Lambda$, in this case when increasing the overall number of dimensions by adding $N$ extra dimensions, then if $N$ is large, the visible dimensions do not have to curve too much, hence a small four dimensional vacuum energy follows.

As we have seen, the stabilization of the potential can be achieved for example by adding a four index field strength taking an expectation values in the visible four dimensional space, as in \cite{DG} and if the other  terms are small, as in the explicit example shown here in the $1/N$ limit, the effective cosmological constant at the minimum can be small with the help of a small additional term, since, as we show explicitly, this results from the interplay of (\ref{e8}), which is small because of the $1/N$ dependence and an additional small $\lambda^2$ term that we  add.

One can now claim that such a procedure to obtain a small vacuum energy does not involve fine tuning. Indeed, the large cosmological constant and the curvature term almost totally cancel out against each other in the large $N$ limit near $\phi=0$, it is only the residual term that needs to be taken care and stabilized with a small vacuum expectation value of a four index field strength.

In this way degrees of freedom, additional to four dimensional gravity, can be responsible for resolving the cosmological constant problem. The effective vacuum energy is reduced by curving the extra dimensions. The higher number of extra dimensions, the more accurate is the cancellation between extra dimensional curvature and the original cosmological constant term (if positive) and appears exact in the limit $N \rightarrow \infty $. For large but finite $N$ a small leftover instability can be taken care by a small vacuum expectation value of a four index field strength. 

This kind of generaliation of four dimensional Einstein gravity in order to confront the cosmological constant problem shares some similar features with for example the "Two Measures Theory" \ct{2measures} , where in four dimensions an additional measure
of integration $\Phi$ independent of the metric in addition to $\sqrt{-g}$, 
so the action used is
\begin{equation}\label{twomeasures}
S = \int L_1 \Phi d^4x +\int L_2 \sqrt{-g} d^4x
\end{equation}
\noindent
then the way that the Two Measures Theory generically resolves the cosmological constant problem is through the dynamical dominance of the measure $\Phi$ (as a consequence of the equations of motion) over the regular measure $\sqrt{-g}$ near the ground state which as a result gets a zero cosmological constant. Notice the resemblance with the $N$ large limit, which is indeed also like enhancing a measure of integration (since in the large number of dimensions limit, we integrate over an infinite dimensional space).

\section*{Acknowledgments}

I  want to thank F. Canfora, A. Giacomini and  A. R. Zerwekh
for very interesting discussions and for informing me about the interesting possibility of considering Kaluza Klein Theories with large number of compact dimensions in the context of the cosmological constant problem and to Aharon Davidson for insights on the $4-D$ Einstein frame picture for studying the Kaluza Klein theories which is key to the present analysis .


\end{document}